\documentclass[runningheads]{llncs}
\usepackage{graphicx}
\usepackage{hyperref}
\usepackage{float}

\usepackage{tikz}
\begin{document}
\title{A Location-Based Global Authorization Method for Underwater Security\thanks{This research was supported by the BRU21 Research and Innovation Program on Digital and Automation Solutions for the Oil and  Gas Industry (www.ntnu.edu/bru21). Pre-submission Comments from Prof. Colin Boyd and Mary Ann Lundteigen are gratefully acknowledged.}}
%
%
\author{Bálint Z. Téglásy\inst{1}\orcidID{0000-0003-0173-608X} \and
Sokratis Katsikas\inst{2,3}\orcidID{0000-0003-2966-9683}}
\authorrunning{B. Téglásy et al.}
%
\institute{NTNU Department of Engineering Cybernetics, O. S. Bragstads Plass 2D, 7034 Trondheim, Norway \and
NTNU Department of Information Security and Communication Technology
\email{sokratis.katsikas@ntnu.no} }
\maketitle              
\begin{abstract}
National or international maritime authorities are used to handle requests for licenses for all kinds of marine activities. These licenses constitute authorizations limited in time and space, but there is no technical security service to check for the authorization of a wide range of marine assets. We have noted secure AIS solutions suitable for more or less constantly internet-connected assets such as ships with satellite connections. The additional constraints posed by underwater autonomous assets, namely less power and connectivity, can be mitigated by using symmetric cryptography. We propose a security service that allows the automatized check of asset authorization status based on large symmetric keys. Key generation can take place at a central authority according to the time and space limitations of a license, i.e. timestamped and geocoded. Our solution harnesses the exceptionally large key size of the RC5 cipher and the standardized encoding of geocells in the Open Location Code system. While we developed and described our solution for offshore underwater use, aerial and terrestrial environments could also make use of it if they are similarly bandwidth constrained or want to rely on quantum resistant and computationally economic symmetric methods.

\end{abstract}
\section{Introduction and Related Work}\label{sec1}

Civilian underwater infrastructure is seen as becoming a target for conflict by authoritative sources on physical security \cite{Focus2021}. Authoritative sources on cybersecurity also see underwater assets being threatened \cite{Risiko2022}. False flag attacks have historically been an even more dominant threat in offshore environments than on land. A concurrent expectation in remote offshore operations is that all control functions are operable from anywhere given the proper security barriers \cite{mikalef2022digital}. Therefore methods for confirming identities (authentication) in the maritime environment are needed \cite{creech_ryan_2003}, and they can be combined with AIS (Automatic Identification System) \cite{goudosis2020secure}. Contracts in civilian offshore operations also define physical access \cite{mikalef2022digital} , but in the case of unmanned assets there are no methods for verifying compliance. An exclusive economic zone (EEZ) ensures exclusive sovereign rights below the surface of the sea \cite{EEZdef}. Due to the advances in subsea technology, the exploitation of EEZs is increasing, while security services that could verify the lawfulness of subsea activities barely exist.

Since communication with the various authorities of a public key infrastructure (PKI) is not feasible in an underwater environment, we have developed and tested authentication methods based on an open physical layer standard (JANUS \cite{Janus}) and symmetric cryptography \cite{AuthNUWAssets} \cite{Branislav}. There are also methods based on symmetric cryptography in standards literature geared towards interoperable underwater communications \cite{Venilia}. Based on these methods, as well as our own previous research \cite{AuthNUWAssets}, it is realistic to assume that underwater communications could be secured much better than today. However, in all of the related work, key generation and distribution \cite{AuthNUWAssets} \cite{Branislav} \cite{Venilia} remains a barely touched upon issue. Underwater assets could receive the necessary keys on the surface to ensure integrity and confidentiality.

However, the methods above consisting of protocols and ciphers leave the question of who gets which pre-shared key(s) or certificate(s) remains unanswered. The expansion with key generation and distribution yields a practicable security service as a cryptosystem. Cryptosystems can be classified as symmetric, asymmetric or hybrid. The advantage of asymmetric cryptosystems lies in better scaleability and guarantees for more security properties, such as non-repudiation. However, devices that are wireless not just in their communication, but also in their power supply rarely harness these benefits. This is because asymmetric cryptosystems are three orders of magnitude more calculation intensive, draining the limited on-board energy. Moreover, asymmetric cryptosystems are likely to be more vulnerable in a postquantum world \cite{IoTPostQuantum}. Underwater wireless devices in particular have a threat landscape dominated by nation-state linked actors \cite{Focus2021} \cite{Risiko2022}, which makes the design of quantum resistant cryptosystems for these assets a timely priority \cite{QuantumDemystifying}.

The difficulty of scalable solutions compound the communication issues, where the scalability of any solution would need to be backed up institutionally. A classification of Symmetric Key Management Schemes for Wireless Sensor Networks provides the different solutions and evaluates their trade-offs \cite{SymmetricKeyClass4Wireless}. Location-aware methods yield themselves for scalable symmetric key management, but are dependant on the spatial deployment patterns \cite{LocationAwareKeyMgmt}. Geocoded key management has been explicitly investigated for asymmetric key management previously \cite{dreyer2020what3words}. We are aware of geocoded symmetric keys being proposed in academic literature before \cite{LocationBasedSymm}. In the latter paper, the implicit assumption is made that latitudes and longitudes are only known to devices that are in those locations. This is not realistic, and can be disproved easily with free and easily accessible tools \cite{LocationOwnSpoof}. Proposals using such a scheme \cite{you2018novel} can not provide much else then a false sense of security. We therefore make a more realistic assumption, namely that the geocoded keys are generated in an offline key ceremony setting and distributed via time-proven Transport Layer Security (TLS) to those and only those devices that are authorized to be present in an area. Any geographic area can be represented as a set of geocells with their corresponding geocodes. The recipients of the keys could verify each other's authorization by looking up the key matching the location to be checked from their individual onboard databases. This way enforcement agencies such as the Coast Guard could carry all keys mapped to their areas of activity and check the authorization of devices found automatically.

Pre-shared keys are needed to secure communications with symmetric cryptography.  Location-based authorization is used anyways to exploit maritime economic resources such as seafood or hydrocarbons. From a security point of view, it would be logical that the authorizations pertaining to exploitation of certain areas are followed with the obligation to authenticate in that area. The long-term keys for said authentication could be attached to the legal authorization issued by the authorities. The keys could be renewed according to rules set by the authority based on the sensitivity of the areas in question, as well as the authentication protocol being used.

Access to the offshore domain is traditionally delimited by location, meaning that the areas defined in maritime law (EEZ, fishery zone, etc.) are subdivided into licenses, quadrants, and blocks that are then licensed to private companies for further use. However, as the offshore economic activities digitalize, many of the digital methods they rely on, such as navigation in general and AIS in particular, remain unsecured. Therefore, the enforcement rules for the above mentioned areas remain plagued by false negatives and false positives of any intrusion detection system. Others have recognized the necessity of including the geographic location data in the key generation process for offshore applications \cite{LocationBasedEncryption}, but have only specified a method with asymmetric encryption.

\section{Methods}

\subsection{Key generation}

A key issue in all security services is how to generate, store and distribute the keys with the identities they are supposed to authenticate or authorize. The key issue is becoming more difficult if we can't or don't want to rely on existing Public Key Infrastructure (PKI). We propose that the geocoded keys are generated by an international organization and distributed to their national partner organization for further distribution as described further below.

\begin{figure}
\centering
\includegraphics[scale=.9]{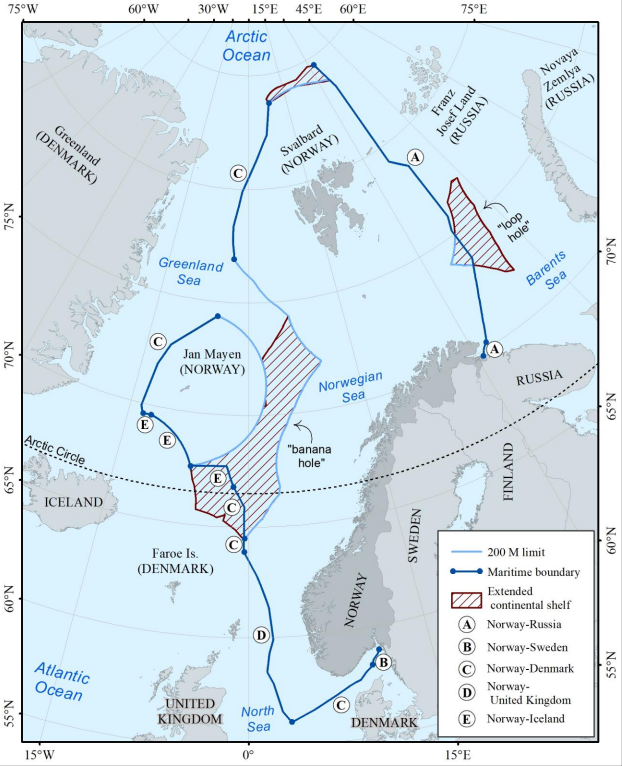}
\caption{\label{fig:NorwayEEZ} The internationally accepted boundaries of Norway's EEZ. Source: \cite{LimitSeas}}
\end{figure}

\begin{figure} 
\tikzset{every picture/.style={line width=0.75pt}} 

\begin{tikzpicture}[x=0.75pt,y=0.75pt,yscale=-1,xscale=1]

\draw    (91.43,109) -- (109.43,109) ;
\draw [shift={(111.43,109)}, rotate = 180] [color={rgb, 255:red, 0; green, 0; blue, 0 }  ][line width=0.75]    (10.93,-3.29) .. controls (6.95,-1.4) and (3.31,-0.3) .. (0,0) .. controls (3.31,0.3) and (6.95,1.4) .. (10.93,3.29)   ;
\draw    (193.43,108) -- (211.43,108) ;
\draw [shift={(213.43,108)}, rotate = 180] [color={rgb, 255:red, 0; green, 0; blue, 0 }  ][line width=0.75]    (10.93,-3.29) .. controls (6.95,-1.4) and (3.31,-0.3) .. (0,0) .. controls (3.31,0.3) and (6.95,1.4) .. (10.93,3.29)   ;
\draw    (257,71) -- (256.48,91) ;
\draw [shift={(256.43,93)}, rotate = 271.49] [color={rgb, 255:red, 0; green, 0; blue, 0 }  ][line width=0.75]    (10.93,-3.29) .. controls (6.95,-1.4) and (3.31,-0.3) .. (0,0) .. controls (3.31,0.3) and (6.95,1.4) .. (10.93,3.29)   ;
\draw    (294.43,38) -- (316.43,38) ;
\draw [shift={(318.43,38)}, rotate = 180] [color={rgb, 255:red, 0; green, 0; blue, 0 }  ][line width=0.75]    (10.93,-3.29) .. controls (6.95,-1.4) and (3.31,-0.3) .. (0,0) .. controls (3.31,0.3) and (6.95,1.4) .. (10.93,3.29)   ;
\draw    (373,71) -- (373.26,88.99) -- (373.37,93) ;
\draw [shift={(373.43,95)}, rotate = 268.37] [color={rgb, 255:red, 0; green, 0; blue, 0 }  ][line width=0.75]    (10.93,-3.29) .. controls (6.95,-1.4) and (3.31,-0.3) .. (0,0) .. controls (3.31,0.3) and (6.95,1.4) .. (10.93,3.29)   ;
\draw    (308.43,204) -- (308.43,231) ;
\draw [shift={(308.43,233)}, rotate = 270] [color={rgb, 255:red, 0; green, 0; blue, 0 }  ][line width=0.75]    (10.93,-3.29) .. controls (6.95,-1.4) and (3.31,-0.3) .. (0,0) .. controls (3.31,0.3) and (6.95,1.4) .. (10.93,3.29)   ;
\draw   (188.83,28) -- (204.03,39) -- (188.83,50) -- (188.83,28) -- cycle (177.43,39) -- (188.83,39) (204.03,28) -- (204.03,50) (204.03,39) -- (215.43,39) ;
\draw   (319.43,151.4) -- (308.43,166.6) -- (297.43,151.4) -- (319.43,151.4) -- cycle (308.43,140) -- (308.43,151.4) (319.43,166.6) -- (297.43,166.6) (308.43,166.6) -- (308.43,178) ;
\draw    (70.43,161) -- (146.65,121.91) ;
\draw [shift={(148.43,121)}, rotate = 152.85] [color={rgb, 255:red, 0; green, 0; blue, 0 }  ][line width=0.75]    (10.93,-3.29) .. controls (6.95,-1.4) and (3.31,-0.3) .. (0,0) .. controls (3.31,0.3) and (6.95,1.4) .. (10.93,3.29)   ;

\draw  [fill={rgb, 255:red, 126; green, 211; blue, 33 }  ,fill opacity=1 ]  (112,95) -- (193,95) -- (193,120) -- (112,120) -- cycle  ;
\draw (115,99) node [anchor=north west][inner sep=0.75pt]   [align=left] {Master key};
\draw  [fill={rgb, 255:red, 126; green, 211; blue, 33 }  ,fill opacity=1 ]  (1,87) -- (91,87) -- (91,133) -- (1,133) -- cycle  ;
\draw (4,91) node [anchor=north west][inner sep=0.75pt]   [align=left] {\begin{minipage}[lt]{58.28pt}\setlength\topsep{0pt}
\begin{center}
Random \\generator(s)
\end{center}

\end{minipage}};
\draw    (1,4) -- (177,4) -- (177,71) -- (1,71) -- cycle  ;
\draw (4,8) node [anchor=north west][inner sep=0.75pt]   [align=left] {Request for authorization\\of activities in an area\\(incl. location, dates)};
\draw  [fill={rgb, 255:red, 126; green, 211; blue, 33 }  ,fill opacity=1 ]  (216,4) -- (294,4) -- (294,71) -- (216,71) -- cycle  ;
\draw (219,8) node [anchor=north west][inner sep=0.75pt]   [align=left] {\begin{minipage}[lt]{50.36pt}\setlength\topsep{0pt}
Determine\\authorized
\begin{center}
geocell(s)
\end{center}

\end{minipage}};
\draw  [fill={rgb, 255:red, 126; green, 211; blue, 33 }  ,fill opacity=1 ]  (214,94) -- (384,94) -- (384,140) -- (214,140) -- cycle  ;
\draw (217,98) node [anchor=north west][inner sep=0.75pt]   [align=left] {\begin{minipage}[lt]{112.73pt}\setlength\topsep{0pt}
\begin{center}
Padding and encryption \\under the master key
\end{center}

\end{minipage}};
\draw  [fill={rgb, 255:red, 74; green, 144; blue, 226 }  ,fill opacity=1 ]  (200,178) -- (411,178) -- (411,203) -- (200,203) -- cycle  ;
\draw (203,182) node [anchor=north west][inner sep=0.75pt]   [align=left] { Geosecured temporary key(s)};
\draw  [fill={rgb, 255:red, 126; green, 211; blue, 33 }  ,fill opacity=1 ]  (319,4) -- (425,4) -- (425,71) -- (319,71) -- cycle  ;
\draw (322,8) node [anchor=north west][inner sep=0.75pt]   [align=left] {\begin{minipage}[lt]{69.6pt}\setlength\topsep{0pt}
\begin{center}
Determine \\authorized\\time interval(s)
\end{center}

\end{minipage}};
\draw  [fill={rgb, 255:red, 74; green, 144; blue, 226 }  ,fill opacity=1 ]  (197,233) -- (411,233) -- (411,279) -- (197,279) -- cycle  ;
\draw (200,237) node [anchor=north west][inner sep=0.75pt]   [align=left] { Key wrapping and distribution\\to surfaced assets e.g. via TLS};
\draw  [fill={rgb, 255:red, 126; green, 211; blue, 33 }  ,fill opacity=1 ]  (1,149) -- (70,149) -- (70,174) -- (1,174) -- cycle  ;
\draw (4,153) node [anchor=north west][inner sep=0.75pt]   [align=left] {\begin{minipage}[lt]{44.1pt}\setlength\topsep{0pt}
\begin{center}
Nonce(s)
\end{center}

\end{minipage}};

\end{tikzpicture}
\caption{The maritime authority generates keys according to the inputs shown here and described below. Green: components of the regular key generation ceremonies. Note that inputs to and outputs from the key ceremonies are protected by data diodes in order to safeguard the master key. Blue: Data and services hosted on the maritime authority server (see also Fig. \ref{fig:distrib})}
\label{fig:method}
\end{figure}
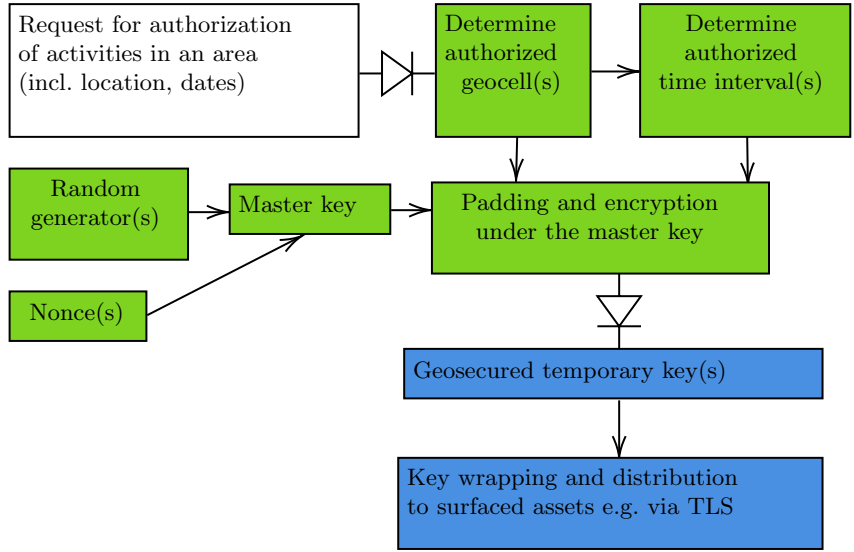
\begin{itemize}
\item Random generator(s) and nonce(s): Different employees of the international authority are responsible for generating the 2040 bit master key collaboratively. The key ceremony could be inspired by those performed at the Internet Assigned Numbers Authority (IANA) every three months to produce and securely store the private portion of Key Signing Keys.
\item Master key: a (k,n) threshold scheme as described by \cite{shamir1979share} will be suitable for the management of this secret. This means that the information representing the master key is accessible when k pieces of out of n are present. n=11 different employees of the central authority are recommended to be responsible for the pieces of the master key, where k=6 of those being present would allow access to the master key.
\item Padding and encryption under the master key: the concatenated geocells and time intervals are first padded to the required length, and then the master key is used to encrypt them. The RC5 cipher is particularly suited to encrypt the geocell and time interval data due to the large key size. This yields the geosecured temporary keys.
\end{itemize}

To illustrate the problem, we take a discrete global grid cell which correspond to a side length of approximately 5.5 km. This grid cell can be geocoded according to 6 base20 digits, as the de facto standard Open Location Code system does. This side length ensures that two modems in the same grid with the described acoustic properties can communicate each other. There are 180 latitude degrees, 360 longitude degrees, and each 1 by 1 degree cell has to be subdivided into a 20*20 subgrid in order to arrive at the sufficiently small side length. Multiplying these factors together gives that there are 25.920.000 such cells. Since the 2040 bit key gives a much larger (\(2^{2040} \approx 1,26e614\)) number of possible keys than the number of possible cells, such a scheme allows for negligible chance of false positives even if all cells are authorised for the exclusive use of different organisations:

Possible keys \(2^{2040}-1 \approx 1,26e614 \gg 25.920.000 \approx 180*360*20*20\) cells

Chosen plaintext attacks are a concern since a lot of the plaintexts (concatenated and padded geocodes and timestamps) might have to be assigned to the same licensees. The large master key size makes chosen plaintext attacks infeasible for deriving it.

Taking into account existing reviews of symmetric key management schemes \cite{bala2013classification}, we believe our proposal is new.  The most similar \cite{LocationAwareKeyMgmt} claims storage and security advantages over previous location-aware solutions, but realizes that attackers can compromise pairwise keys that have been set up using a captured device. While we have no solution for the problem of captured nodes within our scope, we shall also comment on security and storage along those lines:

\begin{enumerate}
\item For fixed assets, the storage of one key might be enough. For authenticating devices in neighbouring cells, one more key corresponding to each of the eight neighbouring cells can be added.
\item If an attacker manages to derive the geocoded keys of an underwater device e.g. through tampering, hostile devices can be placed in the corresponding cells. The sparse population of assets capable of underwater wireless communication minimizes the security consequences in the event of a successful geocoded key capture, and allows for defence in depth to be built up using geographic boundaries.
\item Storing all geocoded keys is possible in less than 7 gigabytes of data. This is a feature allowing federated management of any arbitrary geographic subdivision (as a collection of geocodes) without revealing the master key. Withholding the master key enables delegating the management of any collection of geocodes to another, e.g. national entity, without the danger that this national entity would fabricate keys for geocodes belonging to other entities.
\end{enumerate}

\begin{figure}
\centering
\includegraphics[scale=.31]{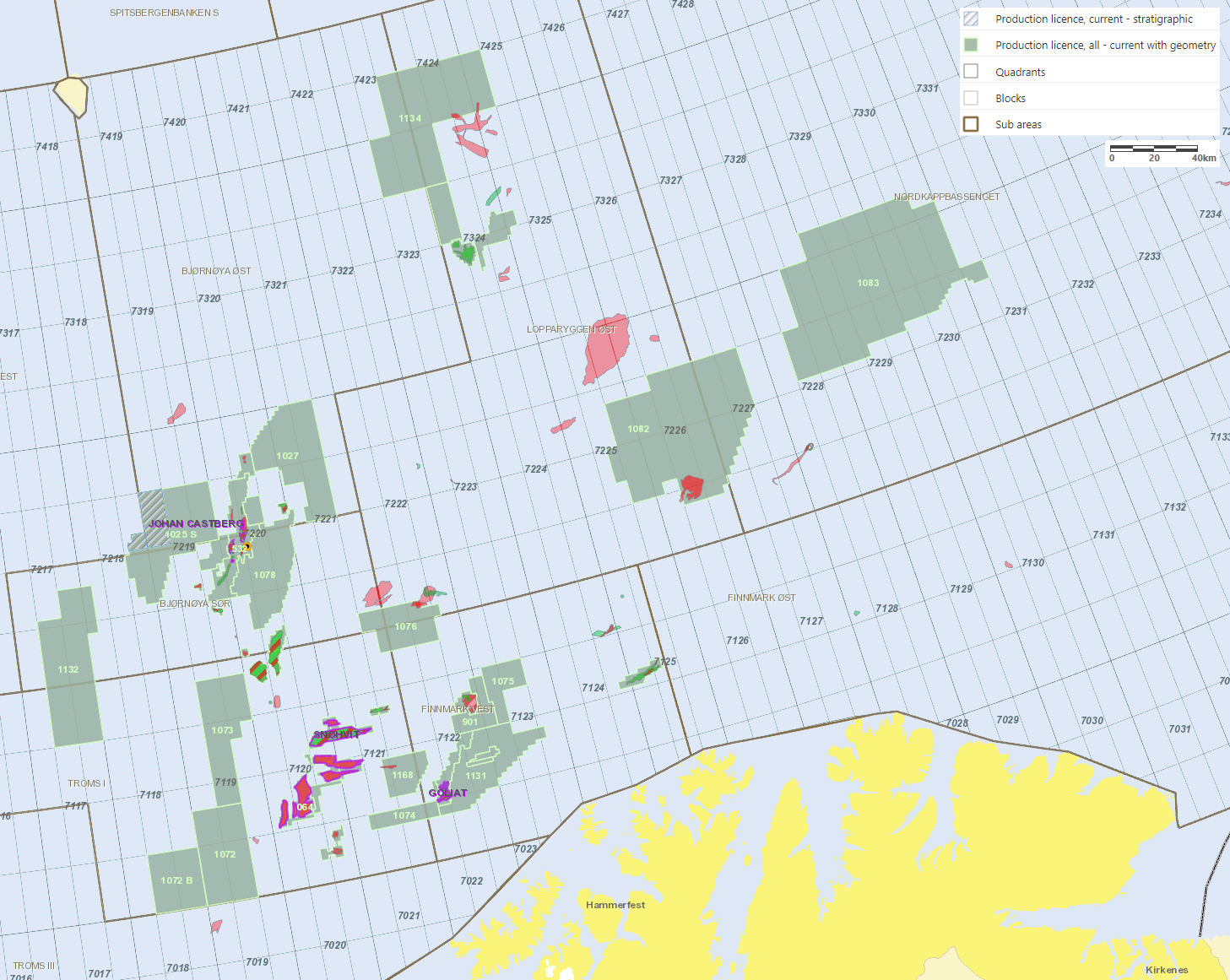}
\caption{\label{fig:barentshavet}Licenses, quadrants and blocks in the Barents sea, excerpt showing part of the area previously shown in Fig. \ref{fig:NorwayEEZ}. Source: \cite{Oljedirektoratet}.}
\end{figure}

\begin{figure}
\centering
\includegraphics[scale=.31]{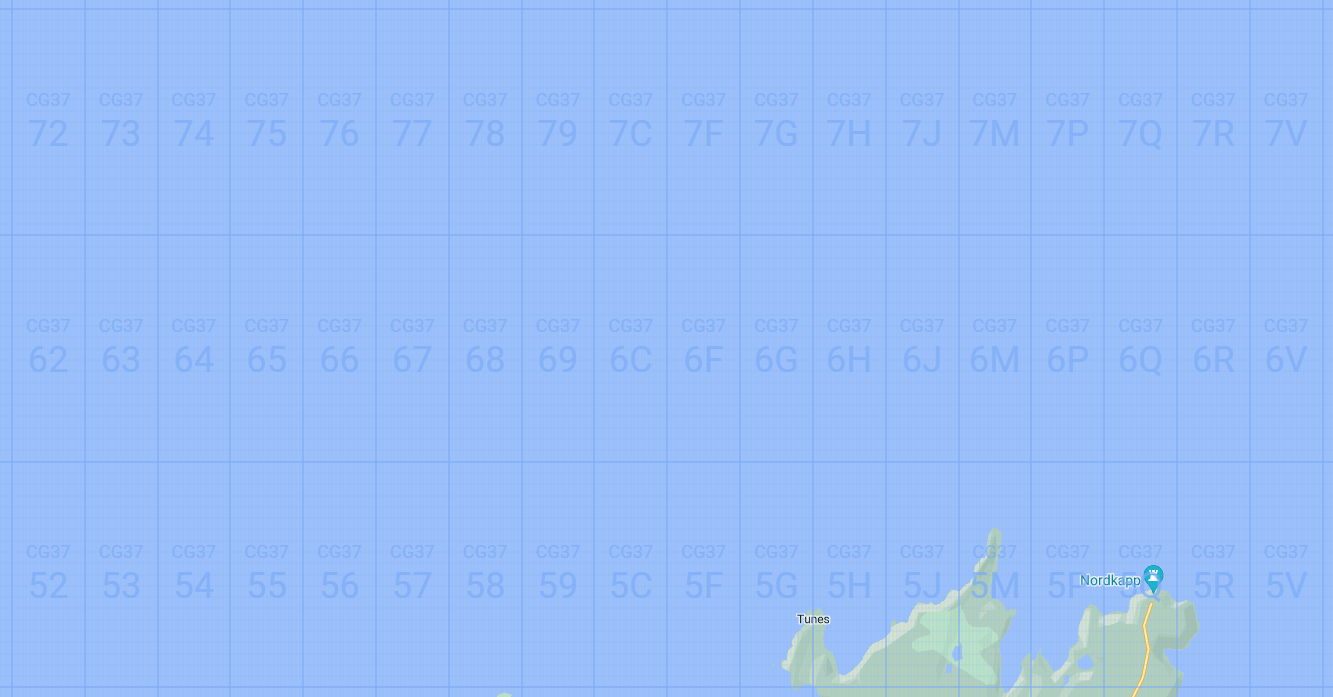}
\caption{\label{fig:nordkapp} Geocodes of the Open Location Code System, excerpt showing part of the area previously shown in Fig. \ref{fig:barentshavet}. Note that the distortion of the square shaped geocells is due to the Mercator map projection at high latitude. Source: \href{plus.codes}{plus.codes}}
\end{figure}

\begin{table}
\caption{Comparison of Symmetric Key Generation Methods for Scalable Systems}
\label{tab:bits}
\centering
\begin{tabular}{|p{1.6cm}|p{7cm}|p{1.8cm}|}
\hline
\textbf{Attributes} & \textbf{Pseudorandom only}& \textbf{Additionally geocoded} \\
\hline
Number of keys assigned& One per organisation worldwide&One per geocode\\
\hline
What is to be proven& Devices belong to the same organization, where the claimed organization has to be assumed. This assumption can be strengthened through associated cleartext communication or through physical layer security, but the false to true negatives and positives ratio is unlikely to reach levels considered secure if the system is to service more than one organization in a geographic area.& The device is in an authorized location\\
\hline
Insider attack consequence& A compromised key means that the whole organisation using that key has lost confidentiality and integrity until surfaced (TLS-mediated) key renewal takes place & Adversarial devices in a geocode gain false authorization until key renewal\\
\hline
\end{tabular}
\end{table}

\subsection{Key distribution}

Key distribution has to happen with the involvement of surface vessels - we hold this to be evident due to the relatively short range of interoperable underwater communication and the spare population of underwater assets. Satellite communication and a public key infrastructure to ensure secure the distribution of data, including keys, should only be assumed for surfaced vessels. To substantiate the assumption, we can look at the long-range identification and tracking (LRIT), an already industrialized secure solution. The root certificate authority for LRIT is operated by the international maritime organization (IMO), but roots of trust for other systems could also be placed at IALA (International Association of Marine Aids to Navigation and Lighthouse Authorities), EMSA (European Maritime Safety Agency) or IHO (International Hydrographic Organization). If a geocoded key generation method is used, it seems useful to co-locate security infrastructure with the Data Centre for Digital Bathymetry (IHO DCDB) such that there are no misunderstandings regarding the correspondance of surface areas to subsea features. This is because bathymetry data includes subsea features linkable to GPS coordinates and therefore geocells. Figure \ref{fig:NorwayEEZ} shows the scale of an area that the international authority has to distribute keys for to a national authority by the example of Norway. Figure \ref{fig:barentshavet} shows how national authorities treat licenses today, an approach that can be secured with the present proposal. Figure \ref{fig:nordkapp} shows the geocells around the north cape of the European continent with their associated geocodes as an example of the view that maritime navigators need to take when challenging or responding to verification requests. Figures \ref{fig:NorwayEEZ}, \ref{fig:barentshavet}, and \ref{fig:nordkapp} include the same area at different scales to facilitate easier understanding.

The final keys distribution should be assigned to existing authorities managing the sovereign rights of a country's EEZ. This could be the Coast Guard in many countries.

Blockchain-Based Certificate Transparency and Revocation Transparency \cite{BlockchainCertificate} might be an attractive choice for an infrastructure that has to provide keys to devices operating in several countries, without relying on any countries’ Certificate Authorities to preserve the integrity of the PKI.

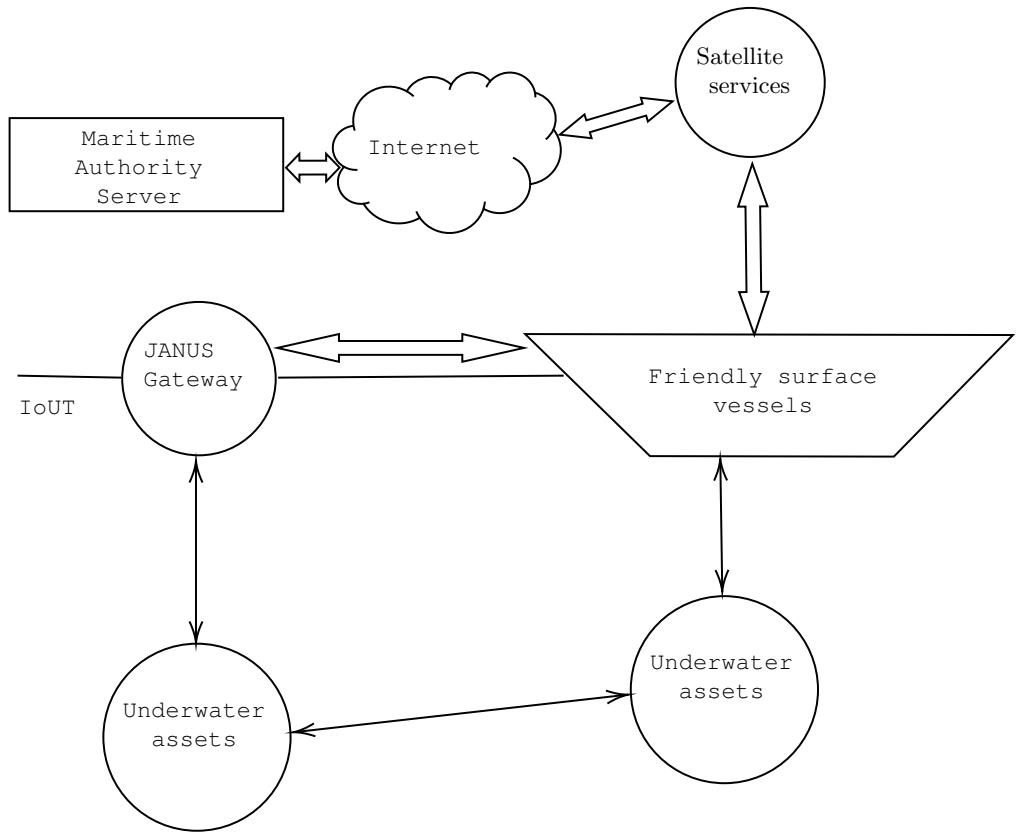
\begin{figure} 
\tikzset{every picture/.style={line width=0.75pt}} 

\begin{tikzpicture}[x=0.75pt,y=0.75pt,yscale=-1,xscale=1]

\draw    (8,187) -- (60.43,188) ;
\draw   (177.47,60.67) .. controls (176.54,54.14) and (179.58,47.67) .. (185.3,44.02) .. controls (191.01,40.36) and (198.4,40.15) .. (204.33,43.49) .. controls (206.43,39.69) and (210.27,37.07) .. (214.7,36.41) .. controls (219.12,35.76) and (223.61,37.15) .. (226.8,40.17) .. controls (228.58,36.73) and (232.1,34.41) .. (236.09,34.05) .. controls (240.08,33.69) and (243.98,35.33) .. (246.41,38.39) .. controls (249.64,34.74) and (254.78,33.2) .. (259.61,34.44) .. controls (264.44,35.68) and (268.08,39.48) .. (268.97,44.19) .. controls (272.93,45.22) and (276.22,47.86) .. (278.01,51.41) .. controls (279.79,54.96) and (279.89,59.09) .. (278.27,62.71) .. controls (282.17,67.58) and (283.08,74.07) .. (280.67,79.76) .. controls (278.25,85.44) and (272.87,89.47) .. (266.54,90.34) .. controls (266.49,95.68) and (263.44,100.58) .. (258.56,103.15) .. controls (253.68,105.72) and (247.74,105.56) .. (243.01,102.73) .. controls (241,109.13) and (235.34,113.83) .. (228.47,114.81) .. controls (221.61,115.79) and (214.77,112.88) .. (210.91,107.33) .. controls (206.18,110.06) and (200.51,110.85) .. (195.17,109.51) .. controls (189.83,108.18) and (185.27,104.82) .. (182.53,100.22) .. controls (177.7,100.76) and (173.02,98.36) .. (170.83,94.2) .. controls (168.64,90.04) and (169.39,85.02) .. (172.72,81.62) .. controls (168.4,79.18) and (166.2,74.35) .. (167.26,69.64) .. controls (168.32,64.94) and (172.4,61.42) .. (177.37,60.92) ; \draw   (172.72,81.62) .. controls (174.75,82.77) and (177.1,83.29) .. (179.45,83.11)(182.53,100.22) .. controls (183.54,100.1) and (184.53,99.86) .. (185.48,99.5)(210.91,107.33) .. controls (210.2,106.3) and (209.6,105.21) .. (209.13,104.06)(243.01,102.73) .. controls (243.38,101.57) and (243.62,100.37) .. (243.72,99.15)(266.54,90.34) .. controls (266.58,84.66) and (263.22,79.46) .. (257.89,76.97)(278.27,62.71) .. controls (277.41,64.65) and (276.09,66.36) .. (274.42,67.73)(268.97,44.19) .. controls (269.11,44.97) and (269.18,45.76) .. (269.17,46.56)(246.41,38.39) .. controls (245.6,39.3) and (244.94,40.31) .. (244.44,41.41)(226.8,40.17) .. controls (226.37,41) and (226.05,41.87) .. (225.84,42.77)(204.33,43.49) .. controls (205.58,44.19) and (206.74,45.04) .. (207.78,46.01)(177.47,60.67) .. controls (177.59,61.57) and (177.79,62.46) .. (178.07,63.33) ;
\draw   (143.42,82.02) -- (150.15,75.02) -- (150.16,78.52) -- (163.66,78.49) -- (163.65,75) -- (170.41,81.97) -- (163.68,88.96) -- (163.67,85.47) -- (150.17,85.5) -- (150.18,88.99) -- cycle ;
\draw    (98,232) -- (98,320) ;
\draw [shift={(98,322)}, rotate = 270] [color={rgb, 255:red, 0; green, 0; blue, 0 }  ][line width=0.75]    (10.93,-3.29) .. controls (6.95,-1.4) and (3.31,-0.3) .. (0,0) .. controls (3.31,0.3) and (6.95,1.4) .. (10.93,3.29)   ;
\draw [shift={(98,230)}, rotate = 90] [color={rgb, 255:red, 0; green, 0; blue, 0 }  ][line width=0.75]    (10.93,-3.29) .. controls (6.95,-1.4) and (3.31,-0.3) .. (0,0) .. controls (3.31,0.3) and (6.95,1.4) .. (10.93,3.29)   ;
\draw    (139.43,188) -- (283.43,187) ;
\draw   (326.98,227.61) -- (264,166) -- (510,166.5) -- (449.98,227.86) -- cycle ;
\draw   (139,173) -- (170.11,166) -- (170.11,169.5) -- (232.32,169.5) -- (232.32,166) -- (263.43,173) -- (232.32,180) -- (232.32,176.5) -- (170.11,176.5) -- (170.11,180) -- cycle ;
\draw    (314.01,348.22) -- (148.99,366.78) ;
\draw [shift={(147,367)}, rotate = 353.59] [color={rgb, 255:red, 0; green, 0; blue, 0 }  ][line width=0.75]    (10.93,-3.29) .. controls (6.95,-1.4) and (3.31,-0.3) .. (0,0) .. controls (3.31,0.3) and (6.95,1.4) .. (10.93,3.29)   ;
\draw [shift={(316,348)}, rotate = 173.59] [color={rgb, 255:red, 0; green, 0; blue, 0 }  ][line width=0.75]    (10.93,-3.29) .. controls (6.95,-1.4) and (3.31,-0.3) .. (0,0) .. controls (3.31,0.3) and (6.95,1.4) .. (10.93,3.29)   ;
\draw    (362.46,231) -- (363.4,294) ;
\draw [shift={(363.43,296)}, rotate = 269.14] [color={rgb, 255:red, 0; green, 0; blue, 0 }  ][line width=0.75]    (10.93,-3.29) .. controls (6.95,-1.4) and (3.31,-0.3) .. (0,0) .. controls (3.31,0.3) and (6.95,1.4) .. (10.93,3.29)   ;
\draw [shift={(362.43,229)}, rotate = 89.14] [color={rgb, 255:red, 0; green, 0; blue, 0 }  ][line width=0.75]    (10.93,-3.29) .. controls (6.95,-1.4) and (3.31,-0.3) .. (0,0) .. controls (3.31,0.3) and (6.95,1.4) .. (10.93,3.29)   ;
\draw   (281.53,65.35) -- (293.96,55.14) -- (294.85,58.14) -- (323.27,49.73) -- (322.38,46.73) -- (338.36,48.53) -- (325.93,58.74) -- (325.04,55.74) -- (296.63,64.15) -- (297.52,67.15) -- cycle ;
\draw   (379.61,166.34) -- (371.96,144.85) -- (375.65,144.81) -- (375.14,101.66) -- (371.44,101.71) -- (378.58,80.05) -- (386.23,101.53) -- (382.53,101.57) -- (383.05,144.72) -- (386.74,144.68) -- cycle ;

\draw (7,198) node [anchor=north west][inner sep=0.75pt]   [align=left] {{\fontfamily{pcr}\selectfont IoUT}};
\draw (183,67) node [anchor=north west][inner sep=0.75pt]   [align=left] {{\fontfamily{pcr}\selectfont Internet}};
\draw    (4,57) -- (142,57) -- (142,104) -- (4,104) -- cycle  ;
\draw (7,61) node [anchor=north west][inner sep=0.75pt]   [align=left] {\begin{minipage}[lt]{91.08pt}\setlength\topsep{0pt}
\begin{center}
{\fontfamily{pcr}\selectfont Maritime Authority}\\{\fontfamily{pcr}\selectfont Server}
\end{center}

\end{minipage}};
\draw    (99.5, 188.5) circle [x radius= 38.71, y radius= 38.71]   ;
\draw (70,169) node [anchor=north west][inner sep=0.75pt]   [align=left] {{\fontfamily{pcr}\selectfont JANUS}\\{\fontfamily{pcr}\selectfont Gateway}};
\draw (384.05,194) node   [align=left] {\begin{minipage}[lt]{106.15pt}\setlength\topsep{0pt}
\begin{center}
{\fontfamily{pcr}\selectfont Friendly surface vessels}
\end{center}

\end{minipage}};
\draw    (364.5, 345.5) circle [x radius= 47.21, y radius= 47.21]   ;
\draw (325,326) node [anchor=north west][inner sep=0.75pt]   [align=left] {\begin{minipage}[lt]{54.87pt}\setlength\topsep{0pt}
\begin{center}
{\fontfamily{pcr}\selectfont Underwater}\\{\fontfamily{pcr}\selectfont assets}
\end{center}

\end{minipage}};
\draw    (98.5, 369.5) circle [x radius= 47.21, y radius= 47.21]   ;
\draw (58.98,350.04) node [anchor=north west][inner sep=0.75pt]  [rotate=-359.94] [align=left] {\begin{minipage}[lt]{54.87pt}\setlength\topsep{0pt}
\begin{center}
{\fontfamily{pcr}\selectfont Underwater}\\{\fontfamily{pcr}\selectfont assets}
\end{center}

\end{minipage}};
\draw    (377.5, 39) circle [x radius= 37.61, y radius= 37.61]   ;
\draw (349,20) node [anchor=north west][inner sep=0.75pt]   [align=left] {\begin{minipage}[lt]{40.13pt}\setlength\topsep{0pt}
Satellite
\begin{center}
services
\end{center}

\end{minipage}};

\end{tikzpicture}
\caption{The proposed hybrid cryptosystem enables the global authorization of underwater assets under the assumption of a secure key distribution above the water.}
\label{fig:distrib}
\end{figure}

\subsection{Rekeying}

Assumptions regarding rekeying for a minimalistic authentication protocol described are in \cite{AuthNUWAssets}, where one baseline packet of the JANUS digital acoustic physical layer protocol is proven to be sufficient. If we follow this to the extent that 29 bit timestamps will be used as a nonce, a rekeying every 60 days is necessary to avoid replay attacks. Introducing additional complexity in the form of key scheduling could expand this rekeying interval without introducing communication overhead in the acoustic domain. The use of initialization vectors would also increase the rekeying interval, albeit at the cost of additional communication overhead.

\section{Results and Discussion}

The proposed methods can potentially be applied for underwater assets in every phase of their life cycle. The applicability hinges upon the ability to host a digital acoustic modem, which can be present already or, depending on the systems under consideration, retrofitted in maintenance outages. For devices that cannot be given access to surface communications regularly, e.g. by surfacing or wiring, key renewal options are limited. They can use session key establishment with other devices that have had access to the surface followed by encrypted long-term keys \cite{AuthNUWAssets}. This method of using other devices as intermediate nodes to receive new keys requires trust. The trust can be rooted in older keys and is therefore not able to repair compromised keys if the adversary is continuously eavesdropping.

\subsection{Assumptions and claims}

Here, we sum up the performance of our system by gathering the assumed inputs and claiming advantages over the state of the art.

We assumed the following:
\begin{enumerate}
\item The devices know their own approximate times and GPS coordinates. They can consequently calculate their own and the neighbouring geocodes.
\item The distribution of centrally generated keys is secure while the devices are surfaced or wired.
\item The devices can store and look up location and time-based keys they've received from the central authority.
\item There exists a protocol by which the keys can be used to check the authorization of another one in the same geocode \cite{AuthNUWAssets}. If the authorization check is successful, they can choose to invest further time and energy into authenticating each other or transmitting data based on the authentication alone.
\end{enumerate}

Symmetric keys like the ones we propose can be used as an authorization step before using another asymmetric scheme to provide the authentication of a unique identifier for the device(s). This would have following advantages:
\begin{enumerate}
\item The repeated use of asymmetric key authentication imposes a computational cost on all involved parties. If any device is entitled to try authenticating based asymmetric keys, this could be a way of denying service by draining resources. In the underwater realm, battery state of charge would be such a resource. Qualifying all devices that need unique identification first by much cheaper symmetric methods mitigates such an attack.
\item Using only asymmetric keys in a threat landscape where post-quantum security is required is prone to disappointments, as recent examples of previously promising candidates show \cite{sidhcastryck2022efficient} \cite{beullens2022breaking}. Even if attacks are combined such that a unique identity is falsely authenticated, there will be a finite number of recipients for the geokey used to authorize the asymmetric authentication. This is a valuable lead for investigations aimed at attributing attacks.
\end{enumerate}

We claim the following advantages over using asymmetric schemes like TLS 1.2 directly underwater:
\begin{enumerate}
\item Much lower computational complexity \(O(n^-3)\) and, as a ramification, quicker execution and lower energy requirements
\item Post-Quantum resistance
\end{enumerate}

We also claim that our proposal to generate symmetric keys by concatenating the outputs of eleven different random generators is more secure then many security services that are accepted as binding by law today. For example a Norwegian Public Key Infrastructure (PKI) substitute called BankID was shown to be vulnerable to insider attacks due to the low information entropy of the key seed \cite{gjosteen2008weaknesses}. The key seed assumed to be generated by the Java SecureRandom class contains 35 bits of entropy in the worst case scenario for the attacker. By running the Java SecureRandom class on eleven compartmentalized devices, preferably with eleven different users responsible for those devices, we claim 385 bits of entropy and therefore mitigation of this type of attack.

\subsection{Potential use cases}

\subsubsection{Stationary}

A fixed asset such as a valve assembly on the seabed might only require the key to the geocode it is in. If the key is compromised by e.g. capturing and tampering by an adversary, the security compromise is confined to that geocell. Such a fixed asset can be a node in any kind of subsea Wireless Sensor Network that are widely discussed in literature.

\subsubsection{Mobile}

A submarine goes on a mission where it is not expected to communicate via radio waves for 40 days. It has an average speed of 25 knots (46 km/h) and therefore a route length of 44160 km. This route could take it through more than 7000 geocells in the EEZs of dozens of countries. In order to enforce sovereign rights, the organisation operating this submarine is required to prove compliance with the authorities of the countries whose EEZ it passes through. It can do so with the use of our method:
1. As a part of mission planning, geocodes that will be passed through are identified. Before submerging, the submarine requests and receives authorization from the relevant authorities of the countries in question. The authorization received from each country includes a table of geocodes, time intervals and their corresponding symmetric keys.
2. The submarine can use the symmetric keys when it is in one of the geocodes without further communication in the electromagnetic spectrum. Based on the physical layer digital acoustic standard JANUS, different methods have been developed that use symmetric keys underwater:
\begin{itemize}
\item The Venilia standard \cite{Venilia} backed by British government agencies. The 256-bit keys that the Tiny Underwater Block (TUB) Cipher therein uses can be derived with minor modifications of our current proposal, such as using the first 256 bits of the RC5 key as the TUB key.
\item We have described a challenge-response identification of friend or foe \cite{AuthNUWAssets}, where bilateral session keys are derived from timestamps. The geocoded temporary keys described in our present method can be used as long-term keys described in our previous paper \cite{AuthNUWAssets}.
The submarine equipped with the geocoded keys may also choose to check the compliance of underwater assets that it could not unequivocally classify based on physical layer acoustic classification that is more traditional in this domain.
\end{itemize}

\subsection{Limitations and generalizations}

The limitations of our method are:
\begin{enumerate}
\item the reliance on secure key distribution through internet protocols like TLS on the surface.
\item inability to verify a unique identity once underwater.
\end{enumerate}

Our method can also be generalized as a kind of key-policy attribute-based encryption (KP-ABE) \cite{goyal2006attribute} for cyber-physical systems (CPS). An important attribute of maritime CPS that is assumed to be known for our method is the location in the form of GPS coordinates. However, there could be many other attributes that can serve to provide fine-grained authorization. In the maritime or aerospace domains, depth or altitude could be an additional extension to the key generation algorithm, such that authorization can be granted in cubical geocodes.

\section{Conclusion} \label{conclusion}

Due to the many limitations of underwater communications, methods for remote security checks of underwater assets are not trivial. We have demonstrated the feasibility of a security service based on geocoded and timestamped symmetric keys. This security service allows the check and consequent enforcement of existing and future offshore licenses. We have identified a range of organizations for providing the services and a concept for the key generation ceremony that maximizes security of the master key.

%
%
\bibliographystyle{splncs04}
\bibliography{mybibliography}

\end{document}